\documentclass[sigconf]{acmart}

\AtBeginDocument{%
  \providecommand\BibTeX{{%
    \normalfont B\kern-0.5em{\scshape i\kern-0.25em b}\kern-0.8em\TeX}}}

\usepackage{verbatim} 
\usepackage{tabularx}
\usepackage{hyperref}
\usepackage{booktabs}
\usepackage{layouts}
\usepackage{subcaption}
\usepackage{graphicx}
\usepackage{picture,xcolor}
\usepackage{dblfloatfix}
\usepackage{comment}
\usepackage{float}
\usepackage{cancel}
\usepackage{footmisc}
\usepackage{todonotes}
\newcommand{\cmmnt}[1]{\ignorespaces}

\definecolor{zz2}{rgb}{0.97,0.96,1.0}

\copyrightyear{2022} 
\acmYear{2022} 
\setcopyright{acmlicensed}\acmConference[CAIN'22]{1st Conference on AI Engineering - Software Engineering for AI}{May 16--24, 2022}{Pittsburgh, PA, USA}
\acmBooktitle{1st Conference on AI Engineering - Software Engineering for AI (CAIN'22), May 16--24, 2022, Pittsburgh, PA, USA}
\acmPrice{15.00}
\acmDOI{10.1145/3522664.3528590}
\acmISBN{978-1-4503-9275-4/22/05}

\begin{document}

\title[Data Smells]{Data Smells: Categories, Causes and Consequences, and Detection of Suspicious Data in AI-based Systems}


\author{Harald Foidl}
\affiliation{
  \institution{University of Innsbruck}
  \country{Austria}
}
\email{harald.foidl@uibk.ac.at}

\author{Michael Felderer}
\affiliation{
  \institution{University of Innsbruck}
  \country{Austria}
}
\affiliation{
  \institution{Blekinge Institute of Technology}
  \country{Sweden}
}
\email{michael.felderer@uibk.ac.at}

\author{Rudolf Ramler}
\affiliation{
  \institution{Software Competence Center Hagenberg GmbH}
  \country{Austria}
  }
\email{rudolf.ramler@scch.at}


\begin{abstract}
High data quality is fundamental for today's AI-based systems. However, although data quality has been an object of research for decades, there is a clear lack of research on \textit{potential} data quality issues (e.g., ambiguous, extraneous values). These kinds of issues are latent in nature and thus often not obvious. Nevertheless, they can be associated with an increased risk of future problems in AI-based systems (e.g., technical debt, data-induced faults). As a counterpart to code smells in software engineering, we refer to such issues as \textit{Data Smells}. This article conceptualizes data smells and elaborates on their causes, consequences, detection, and use in the context of AI-based systems. In addition, a catalogue of 36 data smells divided into three categories (i.e., Believability Smells, Understandability Smells, Consistency Smells) is presented. Moreover, the article outlines tool support for detecting data smells and presents the result of an initial smell detection on more than 240 real-world datasets.
\end{abstract}

\maketitle





\section{Introduction} \label{sec:introduction}
Applications based on artificial intelligence (AI) (e.g., automated driving, predictive maintenance) have grown in popularity over the past decade. However, the resulting AI-based systems pose several challenges \cite{MartinezFernandez_etal2021, Bosch_etal2021}. One of these challenges is their strong data dependency \cite{Schelter_etal2018}. This dependency is caused by data-hungry machine learning (ML) algorithms, typically used in AI-based systems to make intelligent decisions automatically. As a result, poor quality data can lead to abnormal behaviour and false decisions in such systems, resulting in huge monetary losses or, in the worst case, even harming people \cite{Braiek&Khomh2018}.

Recent research (e.g., \cite{Sambasivan_etal2021,Islam_etal2019}), however, suggests that data quality problems are pervasive in AI-based systems. Data quality issues even became one of the main reasons why they suffer badly from technical debt \cite{Bogner_etal2021}.

To improve this situation and thus meet the demand for high data quality in the context of AI-based systems, research in the area of data validation has recently gained significant interest (e.g., \cite{Breck_etal2019, Caveness_etal2020, Redyuk_etal2021, Biessmann_etal2021, Lwakatare_etal2021}). To reliably detect data issues, data validation methods generally require some context-specific constraints, which are usually defined by schemes or rules \cite{Abedjan_etal2016}. However, this declarative and context-dependent nature of data validation combined with the constantly growing amount of data makes data validation a tedious and labor-intensive activity \cite{Schelter_etal2018,Foidl&Felderer2019}.

To address this issue, we previously proposed using \textit{potential} data issues as indicators of latent data quality problems to guide the data validation process in a risk-driven way \cite{Foidl&Felderer2019}. These potential data problems are usually indicated by context-independent, suspicious data values, patterns, and representations, highlighting data that should be prioritised during the validation. We referred to these potential data issues as \textit{data smells} by analogy with code smells.

In the field of software engineering, code smells have become established as indicators of bad design and programming practices that are not faults per se but increase the likelihood of introducing faults in the future \cite{vanEmden&Moonen2012,Fowler_etal1999}. In that sense, code smells are \textit{potential} faults or issues \cite{Santos_etal2018}.

We assert that data smells share several characteristics with code smells. For example, they typically arise due to violated best practices in data handling (e.g., wrong sequence of operations) or data management (e.g., missing data catalogue). Further, they can lead to interpretation issues of software components and impede the evolution and maintenance of AI-based systems. Therefore, we claim that data smells contribute massively to the emergence of technical debt and data-induced bugs within AI-based systems and are thus an increasingly important area of research.

However, although research on data issues is quite mature, it only partly considered such potential data issues. In fact, previous studies (e.g., \cite{Kim_etal2003,Li_etal2011,Josko_etal2019}) mainly focused on actual data errors, lacking a precise definition or categorization of such potential issues. 

This paper aims to address this gap by providing a solid foundation of data smells. In detail, we describe the characteristics of data smells and outline their potential causes, consequences, and use in the context of AI-based systems. We further present a catalogue comprising 36 data smells divided into three categories (Believability, Understandability, and Consistency Data Smells). Moreover, the detection of data smells is discussed and corresponding tool support is presented. Although we consider data smells in the context of AI-based systems in this article, the concept presented is also partially applicable to other data-driven initiatives (e.g., data mining projects).

\begin{figure}[!h]
 \includegraphics[width=0.8\columnwidth,keepaspectratio]{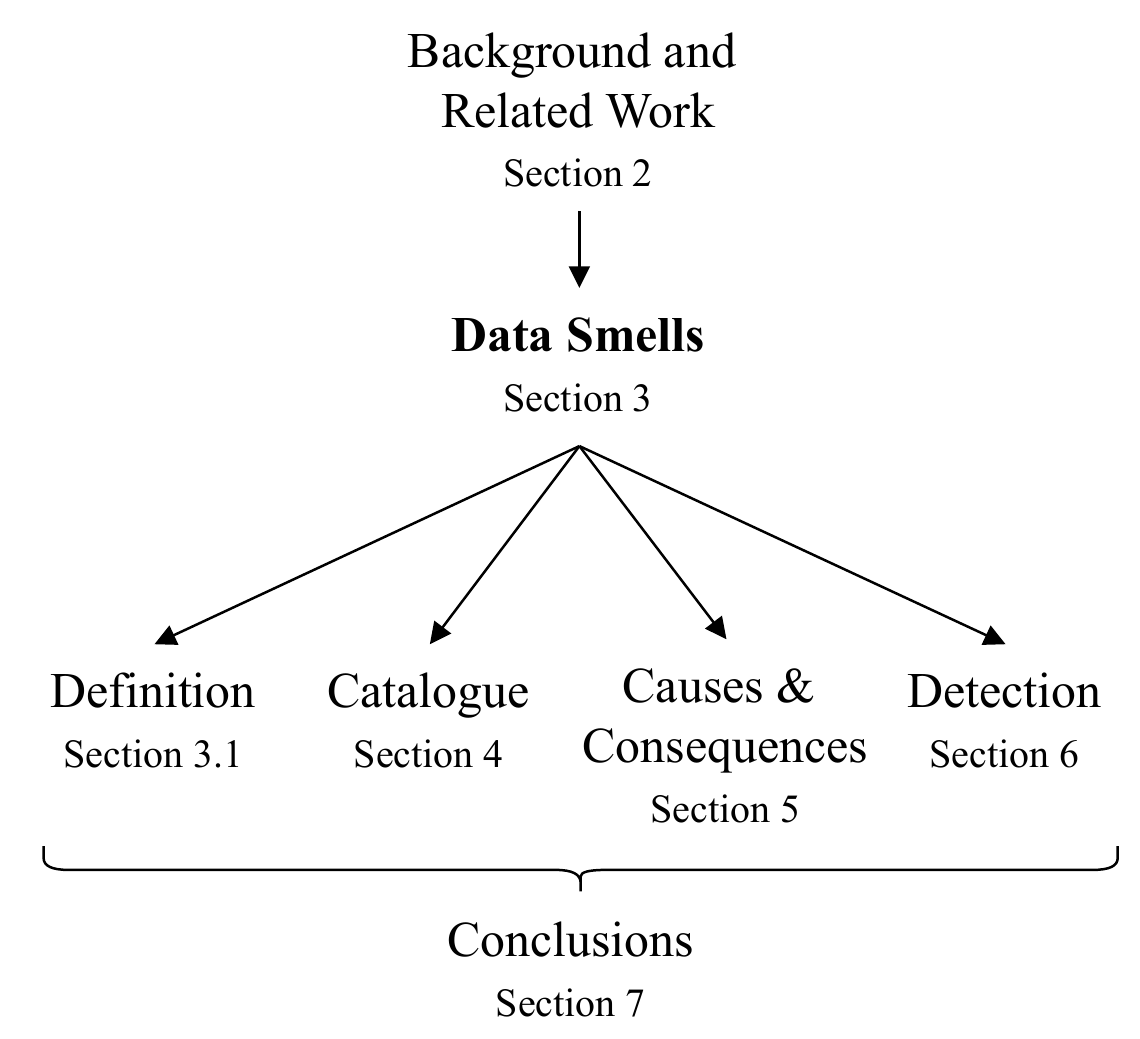}
 \caption{Structure of the article
 \label{fig:gtoc}
}
\end{figure}

The remaining article is structured as outlined in Figure \ref{fig:gtoc}. Section \ref{sec:background} provides background information on the term data smell and discusses related work. The concept of data smells is described in Section \ref{sec:data_smells}. Section \ref{subsec:catalogue} presents the data smell catalogue and Section \ref{sec:causes&conseq} details the causes and consequences of data smells. Afterwards, Section \ref{sec:detection} deals with detecting data smells, thereby presenting detection approaches, corresponding tool support including an experimental evaluation, and use cases. Finally, Section \ref{sec:conclusions} concludes the paper. 

\section{Background} \label{sec:background}
This section first provides a brief overview of the origin and preliminary work on the term data smell in Section \ref{subsec:background_data_smells}. Afterwards, Section \ref{subsec:related_work} presents related work that deals with potential data issues and thus reflects our understanding of data smells.


\subsection{Data Smells} \label{subsec:background_data_smells}
The term "data smells" was first mentioned in the grey literature by Harris in 2014 \cite{Harris2014}. In his article, Harris emphasized the importance of critically examining and questioning data before drawing results and conclusions from them. To this end and based on his software engineering background, he introduced the term data smells for the field of data analysis. Harris mentioned large standard deviations or double-counted records as concrete examples of data smells. However, Harris provided neither a precise definition nor an extensive list of data smells.

In the same year, Iubel \cite{Iubel2014} picked up the term and presented 13 smells for the domain of data journalism on GitHub. However, compared to our work, the presented smells lack a sound categorization and tangible definitions. Nevertheless, the smells Iubel proposed in the category of "unusable data" are generally applicable and not limited to the field of journalism. 

Referring to the academic literature, there is one contribution besides our previous work \cite{Foidl&Felderer2019}, which used the term data smell. Sharma et al. \cite{Sharma_etal2018} introduced the term data smell beside schema and query smell as one kind of database smell. However, they provide a rather vague definition and only illustrate the concept with a single example of a data smell.


\subsection{Related Work} \label{subsec:related_work}
There is a considerable amount of publications addressing issues in data. However, there are plenty of different terms used to refer to data issues, such as dirty data \cite{Kim_etal2003,Li_etal2011}, data error \cite{Abedjan_etal2016}, data defect \cite{Josko_etal2019}, data anomaly \cite{Foorthuis2018,Sukhobok_etal2017} or data quality problem \cite{Oliveira_etal2005}. 
Following, we briefly discuss the most relevant related work that addresses potential data issues, albeit under different terms. See \cite{Batini_etal2016,Ilyas&Chu2019} for an overview of the more general data quality and cleaning concepts.

The most relevant contribution regarding our work was published by Hynes et al. \cite{Hynes_etal2017}. They presented a tool named data linter that aims to detect potential issues in ML training data features. The authors introduced the term data lint as a data- and model-specific inspection rule which identifies data feature representations that are suboptimal to specific ML models. 
Our work differs from their contribution in the following ways. First, unlike data lints, the data smells proposed in this paper describe a more comprehensive concept that also includes the causes and consequences of data quality issues. Second, data smells are generally applicable, while most of the proposed data lints are tailored to ML feature data. Third, we present 36 data smells in contrast to 13 lints proposed by Hynes et al. However, six of the 36 data smells presented are also detectable with the lint detection tool developed by Hynes et al. 

Another field of related work worth to be mentioned has focused on detecting data issues in spreadsheets. Barowy et al. \cite{Barowy_etal2014} presented a technique to investigate whether a data value is erroneous or not. The presented technique is based on the rationale that it is often impossible to know whether a data value has an issue or not. Thus, they propose investigating the impact of a data value on the computation result in spreadsheets. If there is an unusual effect on the result, the data may be erroneous and merit special attention. Similarly, Cunha et al. \cite{Cuha_etal2012} proposed a catalogue of spreadsheet smells as indicators of possible issues in spreadsheet data. 
Although both articles share our understanding of data smells, many of their smells focus on spreadsheet characteristics (e.g., Reference to Empty Cells smell, Standard Deviation smell) and therefore only have limited applicability to other applications.

Other contributions mainly focused on taxonomies of data issues. Although most taxonomies propose proper categorizations of data issues, only a few separate potential data issues in an own category. One of these taxonomies was proposed by Kim et al. \cite{Kim_etal2003}, who grouped such potential data issues as "not missing and not wrong but unusable". Within this category, they list issues such as abbreviations or homonyms. Also, Josko and Wöß \cite{Josko_etal2019} mention "less severe" data defects in their data defect ontology. According to them, such data defects do not threaten the functionality of a system directly but use them to spread to the rest of the system. 

A further work to be mentioned is that of Kasunic \cite{Kasunic_etal2011}. In their work, they define data anomalies to might be erroneous but also as might represent correct data that is caused by unusual but actual circumstances. Although they do not provide further details, it is very close to our understanding of potential data issues.

Sambasivan et al. \cite{Sambasivan_etal2021} recently published a study that, although not explicitly addressing potential data issues, is also related to our work. Simply put, the authors introduced the metaphor of data cascades to frame "compounding events causing negative, downstream effects from data issues" in the context of high-stakes AI systems. Although there are similarities in the origin of data cascades and data smells, both emerge through practices that undervalue data quality, there are significant differences. Data smells describe potential instead of general data issues. Further, data smells are grounded on an engineering perspective, whereas data cascades (e.g., conflicting reward systems) represent a much more high-level concept. 

In sum, the present state of research shows that the idea of potential data issues is not entirely new. However, a sound concept and generally applicable list of data smells is still missing but highly needed. This is precisely where this paper ties in.


\section{Data Smells} \label{sec:data_smells}
This section first defines data smells and describes their main characteristics in Section \ref{subsec:def&char}. Section \ref{subsec:differentation} then discusses how they relate to other types of data quality issues.


\subsection{Definition and Characteristics} \label{subsec:def&char}
In its broadest sense, data smells describe \textit{latent} data quality issues. By latent we mean issues that are present and can cause problems, but are not obvious now. In detail, we define data smells as \textit{context-independent, data value-based indications of latent data quality issues caused by poor practices that may lead to problems in the future}. Following, we describe the main characteristics of data smells concerning their \textit{suspicion}, \textit{context}, \textit{origin} and \textit{consequences}.

\paragraph{Moderate degree of suspicion.}
Data smells are indicated by moderately suspicious data values, patterns and representations that make them challenging to identify. For example, "New York" can be considered smelly because it is semantically unclear whether it refers to the city or the federal state.

\paragraph{Context-Independence.}
Data smells indicate suspicions that are not tied to a specific context. Therefore, they are widely applicable and represent a potential threat for any data-driven system.

\paragraph{Caused by poor practices.}
The emergence of data smells is usually caused by the violation of recommended best practices in data management, software or data engineering (e.g., skip testing of data handling logic). Similarly, data smells are likely to occur due to poor data generation, acquisition or processing (e.g., glue code). Also, a poor intrinsic quality of the data sources from which the data originate (e.g., redundancies in database schema) causes them to arise. 

\paragraph{May cause problems in the future.}
Data smells increase the likelihood of future problems in data processing (e.g., type conversion errors, data misclassifications) and in the evolution of data-driven systems (e.g., enhancement, maintenance). Important in this regard is that the consequences of data smells are usually uncertain and typically delayed in terms of time and location in a data-driven system.


\subsection{Differentiation}\label{subsec:differentation}
To relate data smells to other types of data quality issues, we use the classification of data quality issues proposed by Ge and Helfert \cite{Ge&Helfert2007}. According to them, data quality issues can be divided into context-dependent and context-independent issues. Whereas the latter are independent of the context of use (e.g., missing values, synonyms), the former usually refer to issues that violate contextual rules (e.g., 2.8 as number of siblings). As context-specific knowledge is typically needed to decide if a data value is erroneous or not, we use the term \textit{data error} to refer to context-dependent issues in the remaining paper. 

\begin{figure}[h!]
  \centering 
  \includegraphics[width=\columnwidth,keepaspectratio]{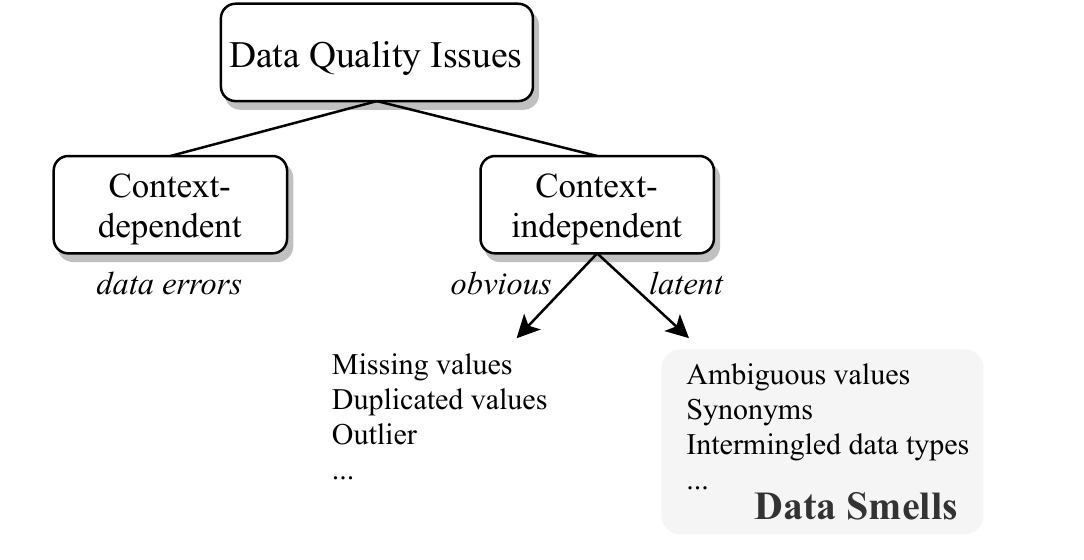} 
  \caption{Classification of data smells in the context of data quality issues}\label{fig:differentation}
\end{figure}

To place data smells into this classification scheme, we propose further subdividing context-independent issues into \textit{obvious} (e.g., missing values, duplicate values, obviously misspelled values) and \textit{latent} issues (e.g., synonyms, ambiguous values) as depicted in Figure \ref{fig:differentation}. In fact, we claim that data smells are a subgroup of context-independent issues representing latent data quality issues. Following, we describe how \textit{obvious} data quality issues and \textit{data errors} mainly differ from data smells.

\paragraph{Obvious data quality issues}
By obvious issues, we mean issues that are indicated by highly suspicious data values, patterns or representations, which are usually recognizable at first glance (e.g., "New Yorc Zity"). Consequently, obvious issues have a higher degree of suspicion compared to data smells which are "only" indicated by moderate suspicions (e.g., "New York"). A further difference with data smells is that obvious issues are typically detectable through basic data profiling techniques (e.g., descriptive statistics). However, obvious data quality issues may have similar consequences as data smells but are much more likely to be discovered and dealt with. 

\paragraph{Data errors}
Data errors are strictly tied to a specific context as assured derivations of the ground truth. This contrasts to data smells which indicate suspicions regardless of the concrete context. In addition, data errors cause inevitable and usually timely consequences, whereas the effects of data smells are uncertain. A further difference to data smells, which typically emerge through poor practices, is that the causes of data errors are manifold (e.g., environmental effects on sensors, violation of domain constraints). \smallskip

Note that both \textit{obvious} and \textit{latent} data quality issues (i.e., data smells) can constitute a data error when set into a concrete context. For instance, the data smell "New York" can represent a data error if the concrete application context only allows country names (e.g., Hungary, Brazil). Thus, although data smells share many properties with code smells, they differ in this respect. Code smells do not represent a bug or an error. Instead, they are symptoms of poor design choices or implementation practices that can only lead to bugs and further issues in the future (e.g., costly maintenance).

\begin{figure*}[h]
       \includegraphics[width=6.0 in, keepaspectratio]{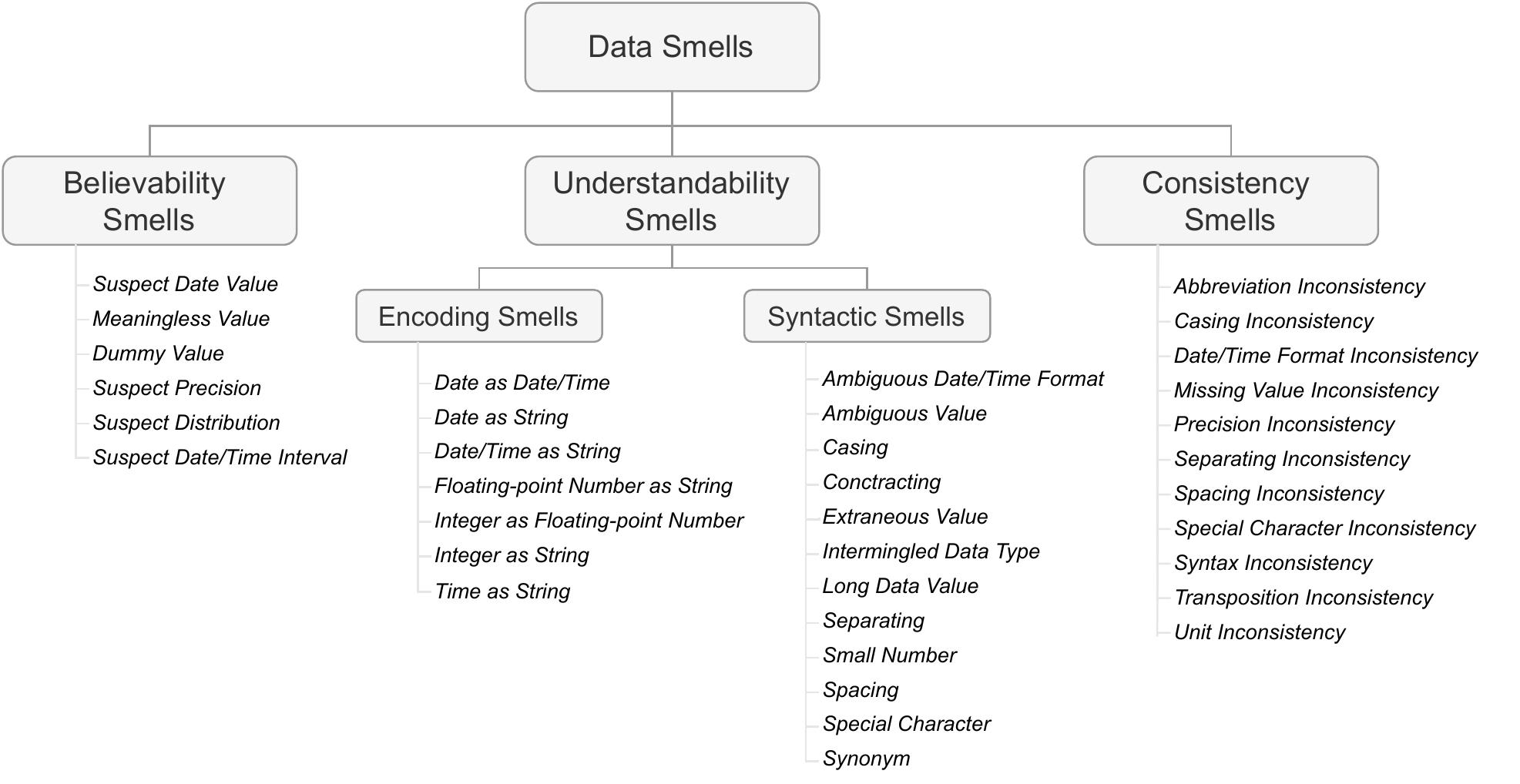}
           \caption{Data smell catalogue}
            \label{fig:catalogue}
\end{figure*}



\section{Catalogue of Data Smells}\label{subsec:catalogue}
In this section, we first outline the research method followed to develop the catalogue in Section \ref{subsec:method}. Subsequently, Section \ref{subsec:categories} presents the catalogue comprising 36 data smells divided into four categories. 

\paragraph{Scope}
We restrict the scope of data smells considered in the catalogue to data of tabular form as this is still the most widely used form of data organisation. In addition, we only consider data smells within one data attribute (i.e., data column). 
Furthermore, we limit the scope of data smells in this paper to three types of data: numerical data (i.e., integer and floating-point values), date/time data (i.e., timestamp values) and text data. These three data types were derived from the most common categories of data used in data-driven systems (i.e., categorical, numerical, time-series and text data).



\subsection{Research Method} \label{subsec:method}
We conducted a multivocal literature review (MLR) \cite{Garousi_etal2019} to develop a sound catalogue of data smells. A MLR was chosen because it allows a deeper insight into practice by additionally considering grey literature on data quality (e.g., blogs, books) and data quality tools (e.g., WinPure\footnote{\label{foot:winpure}\url{https://winpure.com/}}, QuerySurge\footnote{\label{foot:qs}\url{https://www.querysurge.com/}}). 

As a first step, we defined a set of keywords (e.g., "suspicious data", "potential data errors") and conducted a search in Google Scholar and the web search engine Google. We applied an effort bounded stop criterion to limit the search results to a manageable size. Therefore, we relied on the ranking algorithm of Google (Scholar), assuming that the most relevant hits usually appear on the first few pages. In detail, we only checked the first five pages of each keyword's results (i.e., 50 hits) and only continued if the last page contained a relevant hit.  

In total, we checked more than 400 hits and excluded any sources that did not address (at least partly) latent data issues from the remainder of the review. After conducting forward and backward snowballing \cite{Wohlin2014} on the identified sources, we extracted all latent data issues in a spreadsheet\footnote{\label{foot:mlr}\url{https://bit.ly/2WpPauB}}. To ensure that we did not miss relevant issues due to different applied terminology, we additionally extracted common subtle data errors from established data error taxonomies (e.g., \cite{Kim_etal2003,Oliveira_etal2005}). 

As a next step, we reviewed all extracted data issues regarding a moderate degree of suspicion and context-independence. To determine the degree of suspicion, we evaluated the syntactical (i.e., unusual use of characters, formats or data types) as well as the semantical (i.e., implausibility) suspicion of data issues. In addition, we excluded apparent issues (i.e., obvious data quality issues) that are easily identifiable through data profiling techniques. 

After excluding those that did not meet these criteria, we applied an inductive coding approach and assigned descriptive labels to all remaining issues. We then synthesised all labels and derived a list of 36 data smells. The naming of the smells was based on the idea to use the in our opinion most intuitive terms for practitioners. To ensure confirmability of our coding procedure, we conducted a second deductive coding cycle. In doing so, we labelled all extracted data issues again with the corresponding derived data smell(s). 


\subsection{Categories} \label{subsec:categories}
We classified the 36 smells into different categories according to the data quality characteristic they primarily violate. This classification scheme is based on Ganesh et al. \cite{Ganesh_etal2013} who proposed the classification of software design smells according to the violation of object-oriented design principles. Because the terminology on data quality characteristics also differs between authors (e.g., Credibility versus Trustworthiness), we decided to apply the in our opinion most intuitive terms for practitioners. Following, we present the different categories and describe a corresponding smell example. The complete catalogue is depicted in Figure \ref{fig:catalogue}, while a definition and corresponding examples for each smell are available online\footref{foot:mlr}.

\subsubsection{Believability Smells}
Relate to semantically implausible data values. These smells may indicate a low believability of the data values but are usually understandable (i.e., readable) by humans and interpretable by software. 

\smallskip 

\colorbox{zz2}{\begin{minipage}{0.93\linewidth}
 \textbf{\textit{Dummy Value.}}
This smell characterises a situation where a kind of substitute value may be used. As a concrete example, we claim that "999" represents such a smelly value because it is often used to represent missing values and thus is worth investigating. The value probably refers to a common emergency number if the application context is settled around a telephone directory. However, if the value should represent a person's age, it becomes clear that the value possibly indicates a missing entry.
\end{minipage}}

\subsubsection{Understandability Smells} 
Deal with the inappropriate, unusual or ambiguous use of characters, formats or data types. This may lead to problems in interpreting or reading the data values by humans or software, even though the data values are regarded as semantically true. These smells can further be subdivided into Encoding and Syntactic smells.	

\paragraph{Encoding Smells.} 
Represent the inappropriate, unusual or ambiguous use of data types that may lead to data de-/encoding issues. 

\smallskip 

\colorbox{zz2}{\begin{minipage}{0.93\linewidth}
 \textbf{\textit{Integer as String.}}
This smell occurs when an integer is encoded as a string, usually indicated by double quotation marks (e.g., "5"). While some operations can be performed with this data value (e.g., concatenations), others will result in a fault (e.g., additions). 
\end{minipage}}

\paragraph{Syntactic Smells.}
Represent the inappropriate, unusual or ambiguous use of characters or formats that may lead to interpretation issues.

\smallskip 

\colorbox{zz2}{\begin{minipage}{0.93\linewidth}
 \textbf{\textit{Small Number.}}
This smell occurs when data values represent numbers below 1. If several such values are multiplied in subsequent operations, the calculation result can be very small and has many decimal places. As decimal places are often limited manually (e.g., three decimal places) or by the system (i.e., arithmetic underflow), this can lead to incorrect further processing (e.g., 0.02 * 0.02 = 0.000\cancel{4}). 
\end{minipage}}



\subsubsection{Consistency Smells}
Represent the use of inconsistent syntax with respect to data values in partitions of data. 

\smallskip 

\colorbox{zz2}{\begin{minipage}{0.93\linewidth}
 \textbf{\textit{Abbreviation Inconsistency.}}
This smell arises when abbreviations, acronyms or contradictions are not used consistently. For example, the inconsistent use of titles (e.g., Doctor Hill versus Dr. Hill) can lead to malicious results in a natural language processing application.
\end{minipage}}


\section{Causes and Consequences of Data Smells}\label{sec:causes&conseq}
Based on our own experience with industrial data-driven projects and relevant literature (e.g., \cite{Sculley_etal2015,Redgate2017,Breck_etal2019,Dreves_etal2021}), we elaborate on the causes and consequences of data smells in this section. First, Section \ref{subsec:causes} describes the most common causes of data smells. Then, Section \ref{subsec:consequences} illustrates how data smells may affect the correct functioning as well as the development and maintenance of AI-based systems. 


\subsection{Causes}\label{subsec:causes}

\subsubsection{Data management}
Bad practices in data collection or preparation can cause the emergence of data smells (e.g., careless data entry, inconsistent data collection or transformation processes). Further, a lack of documentation (e.g., missing data lineage, no data dictionary) or poor communication between the different actors in the data life-cycle can introduce subtle issues in the data. For example, incomplete metadata can lead to incorrect assumptions about the data by software engineers, resulting in incorrect implemented data processing logic, which causes smelly data. Recent studies outline that poor data practices \cite{Paullada_etal2021,Sambasivan_etal2021} and data-related mismatches \cite{Lewis_etal2021} are pressing issues in the field of AI-based systems. 

\subsubsection{Data handling}
According to a recent study \cite{Yang_etal2021}, data handling code tends to be error-prone and often contains subtle issues. We claim that such poor data handling practices will likely cause data smells. For example, not being explicit when converting a date string (e.g., "2021-01-01") into a datetime object (e.g., pandas.to\_datetime) causes the Date as Date/Time smell (i.e., "2021-01-01 00:00:00") to arise. As the conversion input represents just a date, developers may be unaware that without further declaration, a time suffix (i.e., "00:00:00") is added. 

Such issues often go undetected because of the common programmatic style of method chaining when processing data \cite{Yang_etal2021}. By sequencing multiple data operations (i.e., method chaining), developers cannot see the intermediate processing results and thus identify problems introduced in the data \cite{Wang2021}.

\subsubsection{Data source quality}
Data smells can further arise through a poor intrinsic quality of the data sources from which the data originate. For example, a column name in a relational database is used in different tables but with different data types \cite{Redgate2017}. Accordingly, this can cause data encoding smells when developers retrieve the data as they are not expecting different data types. Often, however, data are already stored smelly in the data sources.


\subsection{Consequences}\label{subsec:consequences}

\subsubsection{Defects and Failures}
The occurrence of data smells can cause interpretation problems of downstream software components in AI-based systems. For example, consider a smell (e.g., Integer as String, Intermingled Data Type) in one of ten thousand data instances of a data attribute to be loaded and processed. When loading data, data processing software libraries typically apply a data type inference algorithm by default (e.g., pandas.read\_csv). As this functionality selects the most flexible data type for a data attribute, a single data smell can lead to a wrong inferred data type \cite{Hynes_etal2017}. Such interpretation problems can cause further incorrect data processing (e.g., type conversion errors) or lead to faults (e.g., arithmetical operations of strings, concatenations of integers) in AI-based systems.

In addition, data smells can also lead to incorrect knowledge generation in an AI-based system. As modern AI-based systems often pursue a continuous learning strategy (e.g., lifelong learning), they typically use serving data as training data. Thus, a well-trained ML model can degrade over time based on smelly serving data continuously used to update the model \cite{Dreves_etal2021}. For example, consider a model that depends on countries as input data. Serving data that becomes smelly, for instance, "US" instead of "us" (i.e., Casing smell), could lead to a wrong result of the model \cite{Polyzotis_etal2018,Dreves_etal2021}.  Due to their high ability to integrate new knowledge (i.e., plasticity), artificial neuronal networks are especially prone to such subtle data issues \cite{Adadi2021}. 

Moreover, the output of one AI-based system is often directly consumed by other systems or even influences its own training data (e.g., direct feedback loops) \cite{Sculley_etal2015, Breck_etal2019}. In such cases, latent data issues can cascade to severe problems over time, causing a gradual regression of the performance of the ML models involved \cite{Caveness_etal2020}.

\subsubsection{Development and Maintainability}
Data smells often require additional data cleaning or preprocessing routines in AI-based systems. For instance, a further code fragment to lower-cases all training data would be needed to address the casing smell mentioned above (i.e., "US" instead "us") \cite{Breck_etal2019}. As such additional data handling code tends to be error-prone \cite{Yang_etal2021} and difficult to understand for developers \cite{Wang2021}, it is supposed to affect the code comprehensibility of AI-based systems. In addition, data handling code is likely to introduce glue code or pipeline jungles in AI-based systems which results in the creation of technical debt over time \cite{Sculley_etal2015}. Consequently, the occurrence of data smells and their corresponding treatment makes AI-based systems harder to understand and maintain and thus increases the risk of introducing further errors. 

Further, data smells make it difficult for humans to understand the data concerned. Accordingly, this can result in wrong assumptions that lead to incorrect implemented functionality or faulty ML models. For example, consider timestamps (e.g., "08:00") represented in 12-hour clock format with missing a.m. or p.m. designators (i.e., Ambiguous Date/Time Format smell). Whereas such time values do not indicate an evident data quality issue, they can be misinterpreted by developers because it is not obvious whether the value refers to eight o'clock in the morning or evening. However, as long as the data describe a casual working day, this smell does not hinder the correct functionality of a ML model. Nevertheless, if shift work is introduced and the model needs to be updated, problems with the correct interpretation of the timestamp "08:00" may arise.

\subsubsection{Real-life Examples} 
Below, we briefly describe how data smells negatively impacted a medical health system and caused problems in COVID-19 software projects.

\paragraph{Oncology Expert Advisor}
The Oncology Expert Advisor (OEA) is an AI-based clinical decision support system developed by IBM \cite{Takahashi_etal2014, Davenport2015}. Based on IBM's cognitive computing system Watson, the main task of OEA is to provide evidence-based therapy recommendations to doctors. Therefore, the system ingests a huge amount of data (e.g., medical literature, practice guidelines, electronic health records). However, during a project with an American cancer center, OEA suffered heavily from latent data quality issues such as acronyms, shorthand phrases or different styles of writing \cite{Ross&Swetlitz2017}. Due to further issues, the project was cancelled in 2016 \cite{Lohr2021}. 


\paragraph{COVID-19 Software Projects}
Consequences of data smells were also reported by an empirical study of COVID-19 software projects \cite{Rahman&Farhana2021}. In this study, the authors investigated the occurrence of bugs in open source software projects that mine and aggregate COVID-19 data. As the second most common bug category, the study identified bugs that occur during processing data (i.e., data bugs). However, many of these data bugs were related to smelly data. For example, a web crawler suffered from a Suspect Date/Time Interval smell \cite{coronadatascraper2020}. In another project, problems arose because the names of some districts in an Indian state differed slightly \cite{covid19indiareact2020}. A different naming for some districts by official sources caused this Ambiguous Value smell to occur. 



\section{Detection of Data Smells}\label{sec:detection}
In this section, we first discuss detecting data smells in the realm of data validation in Section \ref{subsec:preliminaries}. Subsequently, we propose how data smell detection can be approached in Section \ref{subsec:approaches}. Section \ref{subsec:tool_support} then describes the developed tool support for detecting smells. In Section \ref{subsec:evaluation}, we present the conducted experimental evaluation of the tool support. Lastly, Section \ref{subsec:use} discusses the use of data smells in the context of AI-based systems.


\subsection{Preliminaries}\label{subsec:preliminaries}
Data smell detection is closely related to the field of data validation. Validating data has become common practice in AI-based systems as algorithms in these systems strongly rely on the quality of the data fed to them \cite{Biessmann_etal2021}. Basically, data validation in AI-based systems is built upon context-specific schemes and constraints to detect issues in the input data \cite{Breck_etal2018}. Since AI-based systems in production often run continuously in real-time, immediate and straightforward human intervention in case of problems is crucial \cite{Caveness_etal2020}. Thus, data validation systems are typically tailored to provide reliable, high-precision alerts about actual data errors \cite{Breck_etal2019,Dreves_etal2021}. Notable solutions for validating data in AI-based systems are Amazon's Deequ \cite{Schelter_etal2018, Schelter_etal2018b} or Google's TensorFlow Data Validation (TFDV) \cite{Caveness_etal2020}.

Although integrating data smell detection into data validation systems seems evident at first glance, we propose to decouple it for the following reasons.

First, detecting data smells will likely produce many alerts about potential issues. On the one hand, this vast amount of alerts is caused by the more generally specified detection methods needed to ensure context-independence detection. On the other hand, the stringent definition of detection thresholds needed because of the only moderate suspicion of smelly data also contributes to a large number of alerts. This large number of alerts is problematic because determining whether a detected smell really causes problems in AI-based systems requires substantial human effort. In fact, a smell must be set into the concrete application context and requires a detailed examination of a system's downstream components and applications. Thus, detecting smelly data during validating data contrasts with the aim of data validation systems to provide precise, immediate and actionable notifications about actual errors in the data \cite{Polyzotis&Zaharia2021}.

Second, applying data quality checks on continuously arriving data may affect the performance of AI-based systems (i.e., latency) \cite{Merino_etal2020}. Therefore, the amount and stringency of the detection methods used in data validation systems must be treated with caution. Accordingly, this also contradicts the integration of data smell detection into data validation systems.

In summary, smelly data are basically not data errors and are therefore not suitable to be detected in data validation systems. Since data smells typically arise through poor practices, the primary goal should not be to detect and clean them in real-time but to identify and resolve their root causes. This is closely related to the practice of refactoring to eliminate code smells in software engineering.




\subsection{Detection Approach}\label{subsec:approaches}
As outlined in the previous section, detecting data smells as part of data validation modules in AI-based systems would come with several drawbacks. We thus propose to detect data smells in an offline manner without affecting systems running in production. To deal with the presumably large number of smell alerts and to vary the level of suspicion to be flagged, we further introduce two metrics.

\paragraph{Data Smell Strength} 
This metric indicates the likelihood that a data value (i.e., data instance) or pattern (i.e., data partition) is treated as suspicious, and a smell is raised. Basically, this metric implies the concrete thresholds and/or hyperparameters of the individual detection methods. For example, the number of contiguous characters required to detect a Long Data Value smell. 

\paragraph{Data Smell Density} 
This metric describes the relative number of detected smells of a data attribute (i.e., data column). Therefore, this metric can be used to focus on data attributes with a high density of smells. For example, a data attribute can only be considered smelly (i.e., \textit{Smelly Data Attribute}) if at least 10 percent of its data instances represent a data smell.


\subsection{Tool Support}\label{subsec:tool_support}
To operationalize the data smell detection, we checked several data quality-related tools (e.g., OpenRefine\footnote{\url{https://openrefine.org/}}, WinPure\footref{foot:winpure}) 
as well as several data validation tools and libraries, such as Deequ \cite{Schelter_etal2018}, TFDV \cite{Caveness_etal2020}, Data Sentinel \cite{Swami_etal2020}, MobyDQ \cite{MobyDQ2021}, Data Quality Toolkit \cite{Gupta_etal2021}, Data Quality Advisor \cite{Shrivastava_etal2019} and Great Expectations \cite{GreatExpectations2021} for their ability to detect smelly data. Most of the investigated tools have a particular focus (e.g., outlier detection), are not publicly available (e.g., \cite{Swami_etal2020}), or come with a limited, fixed set of built-in data checks (e.g., \cite{Gupta_etal2021}). Thus, we did not find any tool or library that is able to detect most of the proposed data smells without adaptions. 
We therefore decided to develop an own solution to enable automated data smell detection.

\begin{figure}[h] 

\subfloat[Rule-based detection]{%
 \includegraphics[width=\columnwidth,keepaspectratio]{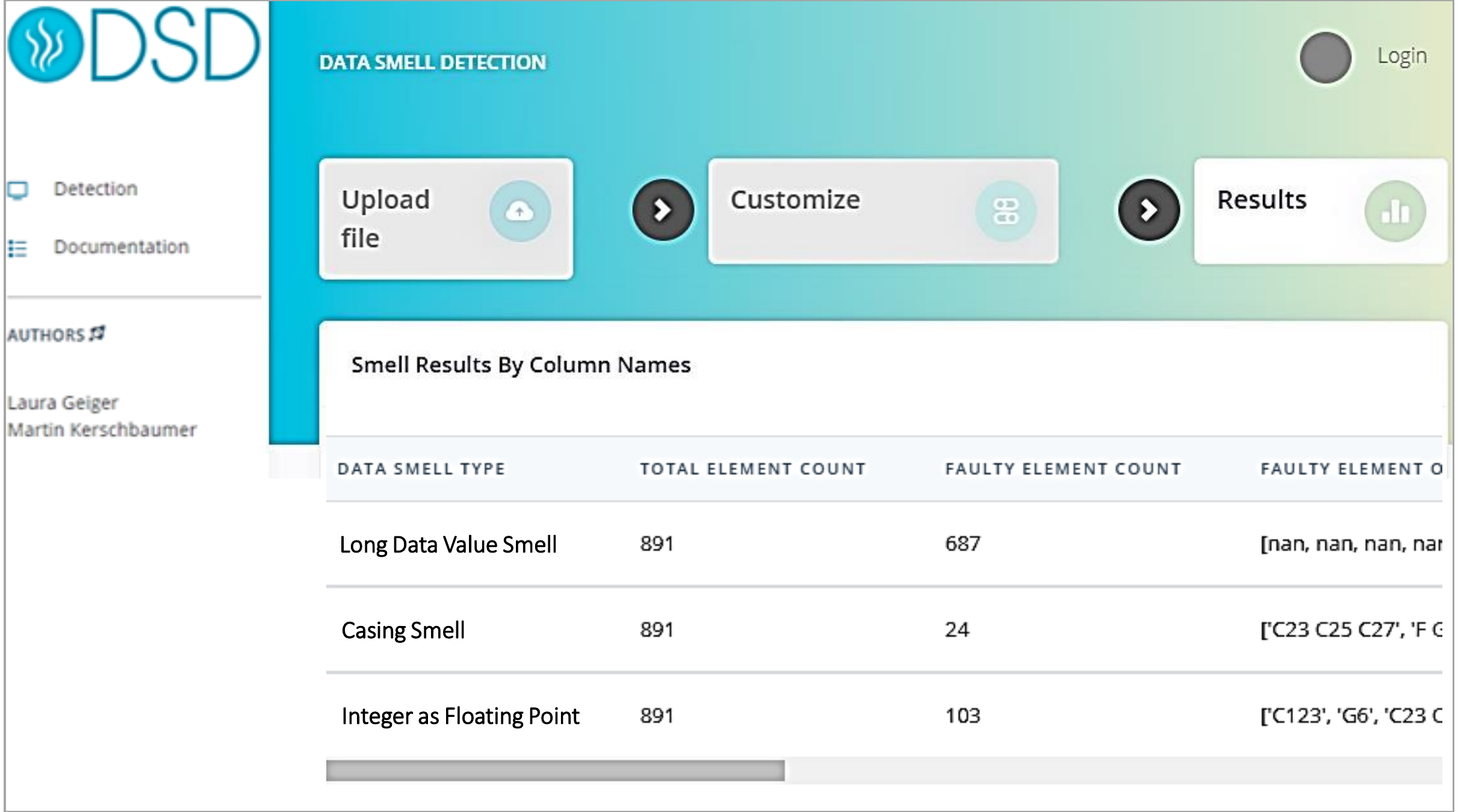}
 \label{fig:tool_support_rb}
}

\subfloat[ML-based detection]{%
   \includegraphics[width=\columnwidth,keepaspectratio]{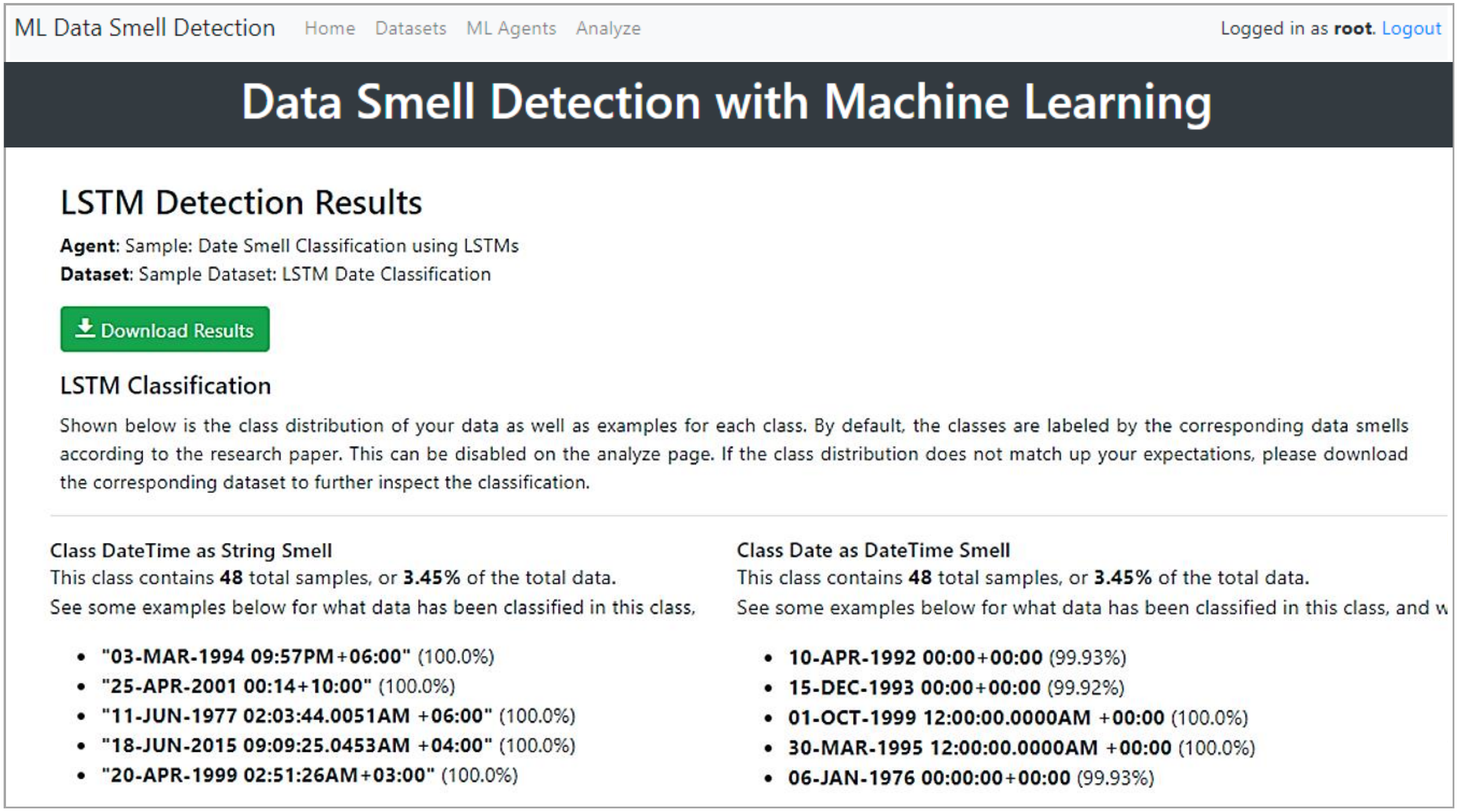}
    \label{fig:tool_support_ml}
}
\caption{Tool support}\label{fig:tool_support}

\end{figure} 

After reviewing each defined smell from the catalogue, we concluded that while some data smells (e.g., Long Data Value smell) are rather easy to spot using regular expressions, others (e.g., Synonym smell) require more advanced techniques to be reliably detected. Therefore, we propose to detect data smells using a \textit{rule-based} and a \textit{ML-based approach}. Traditional data validation techniques (e.g., range and data type checks) can be adapted and used to implement rule-based smell detection. For realizing a learning-based approach, techniques from natural language processing (e.g., recurrent neural networks) or anomaly detection (e.g., autoencoders) are worthwhile candidates. 

Following, we present the two tools developed based on the outlined detection approaches. Several detection methods implemented in the tools were calibrated and trained with many publicly available datasets to ensure general applicability.

\subsubsection{Rule-based Detection}
The first tool\footnote{\url{https://github.com/mkerschbaumer/rb-data-smell-detection}} is based on the open-source data validation tool \textit{Great Expectations\footnote{\url{https://greatexpectations.io}}} and focuses on a rule-based smell detection. Great Expectations was chosen due to its ability to be easily extensible and thus suitable for implementing data smell-specific detections. By adjusting so called expectations (i.e., assertions about data) provided by Great Expectations, the tool is able to detect smells such as Long Data Value, Casing, Integer as String, Floating Point Number as String or Integer as Floating Point Number. The tool provides a graphical user interface where CSV files can be uploaded. Further, the degree of suspicion to be flagged can be easily chosen based on predefined settings. However, it is also possible to define each parameter of a detection method individually.


\subsubsection{Machine learning-based Detection}
The second tool\footnote{\url{https://github.com/georg-wenzel/ml-data-smell-detection}} aims to detect data smells based on several ML algorithms. For example, we used the Word2Vec algorithms implemented in the Python library Gensim to realise the Synonym smell detection. Further, several inconsistency smells (e.g., Casing and Spacing Inconsistency) were implemented by using the Python library Dedupe. Neuronal networks (e.g., long short-term memory, autoencoders) for detecting smells such as Ambiguous Date/Time Format or Date/Time Format Inconsistency were realised with the library Keras. The developed tool also comes with a graphical user interface and accepts CSV files. Most of the implemented ML models were trained for general applicability but also allow to be retrained on user-specific data. A screenshot of detection results of both tools is depicted in Figure \ref{fig:tool_support}.



\subsection{Experimental Evaluation}\label{subsec:evaluation}
We applied several data smell detection methods on 246 Kaggle datasets to test and evaluate the tool support.

\subsubsection{Setting}
\begin{figure}[H]
 \includegraphics[width=\columnwidth,keepaspectratio]{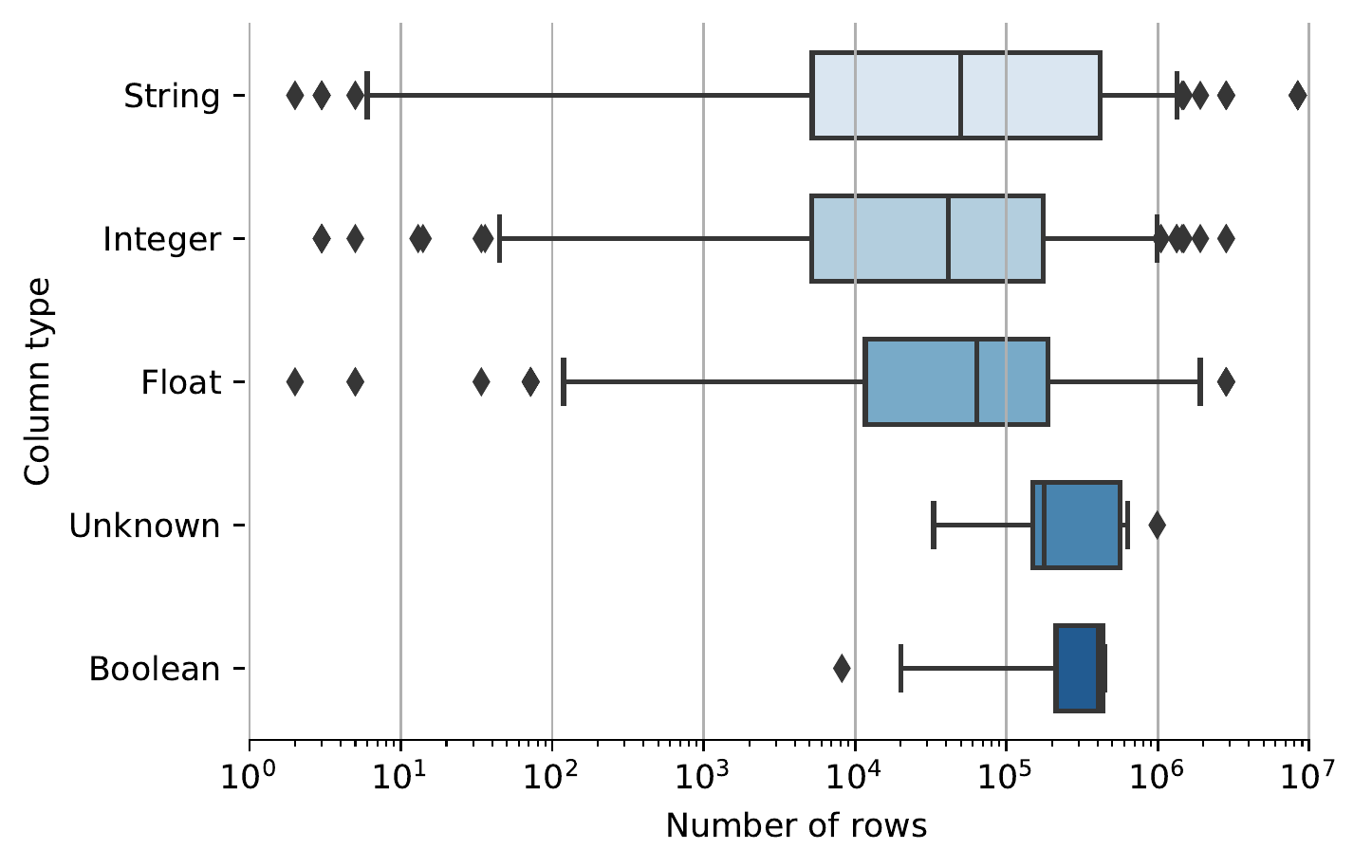}
 \caption{Boxplot of number of rows grouped by column type
 \label{fig:boxplot}
}
\end{figure}

All datasets were randomly selected and are licensed under the Creative Commons CC0. To conduct the smell detection, the datasets were first downloaded and further processed with Python scripts to be analysable by the tool support. In total, the datasets resulted in more than 2,000 columns and more than 42 billion rows to be analysed. The majority of the columns were of type String (i.e., 1,130). Figure \ref{fig:boxplot} visualizes the number of rows for each column grouped by its type. Although the number of rows varies widely (i.e., 2 to 8,405,079), 50 percent of each considered column type contained more than 41,500 rows.

\subsubsection{Results}
\begin{figure}[H]
\includegraphics[width=\columnwidth,keepaspectratio]{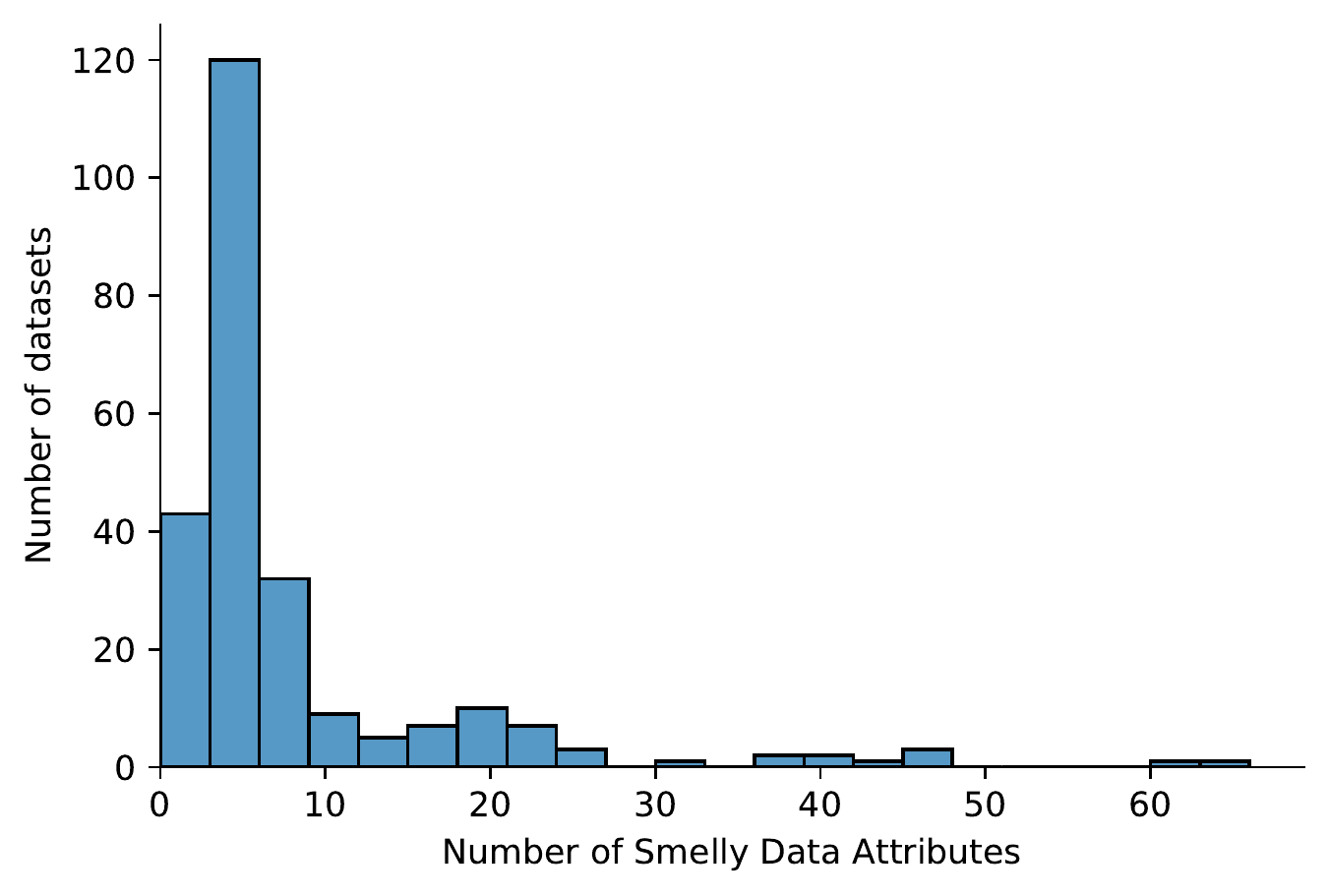}
\caption{Histogram of smelly data attributes across all datasets}\label{fig:histogram}
\end{figure}

We analysed the number of Smelly Data Attributes for each dataset to get an initial impression on the occurrence of smells in real-world data. An attribute (i.e., column) was considered smelly if at least one smell was detected. Figure \ref{fig:histogram} shows the corresponding frequency distribution. 

It is apparent from this figure that most of the datasets (i.e., 120) contained three to five smelly attributes. Only 45 datasets were detected with more than 10 smelly attributes. In contrast, three datasets had no data smells. These initial observations suggest that the number of detected smelly attributes seems to be at a manageable size for real-world data.

\begin{figure*}[!t]
\centering
\subfloat[Ambiguous Date/Time Format smell]{%
 \includegraphics[width=0.32\linewidth,keepaspectratio]{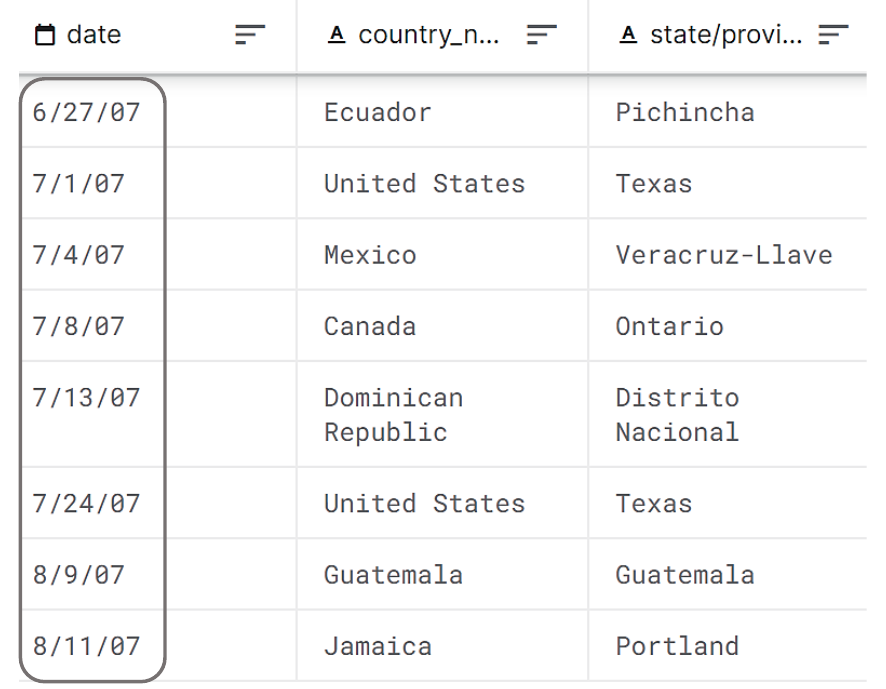}
 \label{fig:ADT1}
}
\subfloat[Ambiguous Date/Time Format smell]{%
   \includegraphics[width=0.32\linewidth,keepaspectratio]{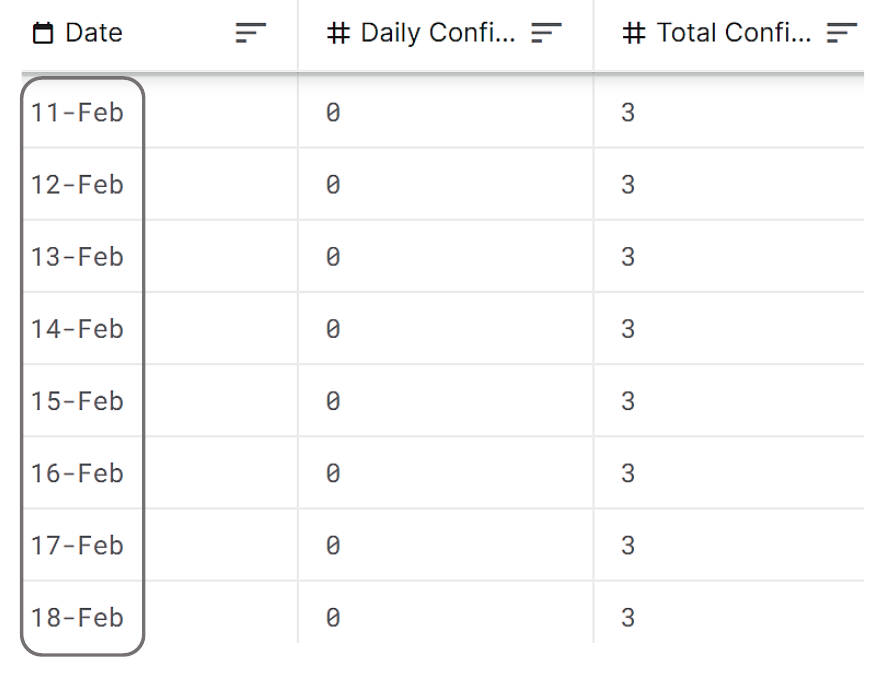}
    \label{fig:ADT2}
}
\subfloat[Date as Date/Time smell]{%
   \includegraphics[width=0.32\linewidth,keepaspectratio]{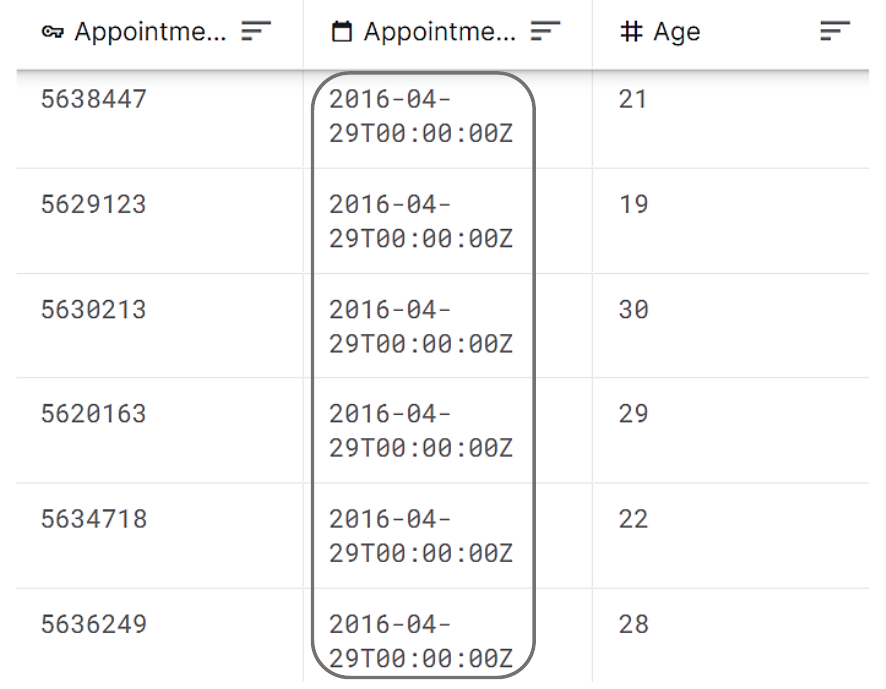}
    \label{fig:DD}
}
\caption{Data smells in real-world data}\label{fig:eva-real_world}
\end{figure*}

\paragraph{Exemplary Smells}
Figure \ref{fig:eva-real_world} shows an excerpt of detected smells in real-world data. On the left-hand side, Subfigure \ref{fig:ADT1} shows an example of the Ambiguous Date/Time Format smell\footnote{\url{https://www.kaggle.com/nasa/landslide-events}}. In fact, the date is represented in short format, which may lead to interpretation problems as it is unclear which date format is applied. Subfigure \ref{fig:ADT2} also depicts an example of the Ambiguous Date/Time Format smell. The corresponding COVID-19 dataset\footnote{\url{https://www.kaggle.com/ravichaubey1506/covid19-india}} represents case data but omits the year in the date column. Thus, without further information (e.g., metadata), it can lead to wrong assumptions about the corresponding year. In Subfigure \ref{fig:DD}, an example of the Date as Date/Time smell\footnote{\url{https://www.kaggle.com/joniarroba/noshowappointments}} is shown. The corresponding column represents the day of medical appointments with a time suffix (i.e., "00:00:00"). Thus, systems consuming these data may encounter problems in processing these data correctly.


\subsection{Use}\label{subsec:use}
Detecting data smells is especially useful in the engineering phase of AI-based systems. A data validation system has typically not yet been implemented in this phase, and data quality assurance is thus often neglected. However, data smell detection can be applied with little effort due to the context-independent nature of smelly data. For instance, by checking training data for smells, potentially problematic data sources can be identified before they are later used in production. Furthermore, data smells can guide data validation efforts in environments with many different data streams. For example, since validating thousands of data signals is impossible, data engineers can consider the smell density (i.e., Data Smell Density), besides other aspects (e.g., feature importance), when deciding which data fractions to validate. Additionally, data smells can uncover subtle issues in the data handling logic or software code that passed traditional quality assurance processes and thus would creep into the deployed system.

Further, checking data on the occurrence of smells is also useful on systems running in production. In fact, regularly detecting smells on archived serving data can identify data issues that were not caught by a system's data validation components. Thus, root causes of these smells can be identified and fixed before the smelly data may cause problems over time (e.g., model degradation, affecting other systems). If the root causes cannot be identified or resolved, data validation components can at least be adjusted (e.g., strengthen their constraints) to catch these smelly data. 

In summary, detecting data smells effectively reduces technical debt and increases the quality of data in AI-based systems.


\section{Conclusions} \label{sec:conclusions}

The main aim of this article was twofold. First, we have addressed the lack of research on latent data quality issues by conceptualizing them in the form of data smells. Therefore, we have presented a sound definition, the main characteristics, and a catalogue of data smells by analogy to the concept of code smells in software engineering. 

Second, we have highlighted the importance of the often neglected yet pervasive, smelly data in AI-based systems. Therefore, we have explicitly elaborated on the causes and consequences of data smells in such systems. In detail, we have shown that bad practices (i.e., data management and handling) in engineering AI-based systems causes smelly data to arise. We further have demonstrated how data smells can negatively impact the correct functioning as well as the development and maintenance of AI-based systems. Two presented cases (i.e., IBM's Oncology Expert Advisor, COVID-19 software projects) illustrated the consequences and currentness of smelly data in real-world projects. 

In addition, we have discussed the detection of data smells and proposed a rule-based as well as a ML-based detection approach decoupled from traditional data validation efforts in AI-based systems. We implemented both detection approaches as tool support and conducted an initial experimental evaluation on 246 Kaggle datasets. The evaluation indicated that the number of detected data smells seems to be at a manageable size for real-world data. Lastly, we have proposed application scenarios of detecting smells to provide a solid baseline for future research.

In fact, we intend to identify less and more influential smells regarding their impact on AI-based systems in the future. For this purpose, we plan to apply a framework proposed by Schelter et al. \cite{Schelter_etal2021} (i.e., JENGA) which is able to study the impact of data issues on ML models.
Moreover, future research should investigate whether using data smell detection as part of testing software components to spot subtle implementation errors is beneficial. On a wider level, research should also focus on determining data smells for specific domains (e.g., finance, industry, medicine). There are already contributions to build on, such as a publication by Ehrlinger et al. \cite{Ehrlinger_etal2018}, which investigates patterns of missing industrial data.

\begin{acks}
The research reported in this paper has been partly funded by the Federal Ministry for Climate Action, Environment, Energy, Mobility, Innovation and Technology (BMK) and the Federal Ministry for Digital and Economic Affairs (BMDW) as well as the State of Upper Austria in the frame of the COMET - Competence Centers for Excellent Technologies Programme (FFG-Nr. 865891) and the project ConTest (FFG-Nr. 888127) managed by Austrian Research Promotion
Agency FFG. We also thank Martin Kerschbaumer, Georg Wenzel and Laura Geiger for their contribution in the implementation of the tool support.
\end{acks}

\bibliographystyle{ACM-Reference-Format}
\bibliography{literature}

\end{document}